\documentclass[aps,pra,twocolumn,superscriptaddress]{revtex4-1}
\usepackage{amsfonts,amssymb}
\usepackage{amsmath,times}
\usepackage{graphicx}
\usepackage{hyperref}    
\usepackage{color}

\newcommand{\uk}{u_{\mathbf{k}}}
\newcommand{\vk}{v_{\mathbf{k}}}

\begin{document}
\title{Strong pairing in two dimensions: Pseudogaps, domes, and other implications}
\author{Xiaoyu Wang}
\affiliation{James Franck Institute, University of Chicago, Chicago, Illinois 60637, USA}
\author{Qijin Chen}
 \affiliation{Shanghai Branch, National Laboratory for Physical Sciences at Microscale and Department of Modern Physics, University
  of Science and Technology of China, Shanghai 201315, China}
\affiliation{Zhejiang Institute of Modern Physics and Department of Physics, Zhejiang University, Hangzhou, Zhejiang 310027, China}
\author{K. Levin}
\affiliation{James Franck Institute, University of Chicago, Chicago, Illinois 60637, USA}

\begin{abstract}
  This paper addresses the transition from the normal to the
  superfluid state in strongly correlated two dimensional fermionic
  superconductors and Fermi gases.  We arrive at the
  Berezinskii-Kosterlitz-Thouless (BKT) temperature $T_{\text{BKT}}$
  as a function of \emph{attractive} pairing strength by associating
  it with the onset of ``quasi-condensation" in the normal phase.  Our
  approach builds on a criterion for determining the BKT transition
  temperature for atomic gases which is based on a well established
  Quantum Monte Carlo analysis of the phase space density.  This
  latter quantity, when derived from BCS-BEC crossover theory for
  fermions, leads to non-monotonic behavior for $T_{\text{BKT}}$ as a
  function of the attractive interaction or inverse scattering length.
  In Fermi gases, this implies a robust superconducting dome followed
  by a long tail from the flat BEC asymptote, rather similar to what
  is observed experimentally. For lattice systems we find that
  $T_{\text{BKT}}$ has an absolute maximum of the order of $0.1 E_F$.
  We discuss how our results compare with those derived from the
  Nelson Kosterlitz criterion based on the mean field superfluid
  density and the approach to the transition from below. While there
  is agreement in the strict mean-field BCS regime at weak coupling,
  we find
  that
  at moderate pairing strength
  bosonic excitations cause a substantial increase in $T_\text{BKT}$
  followed by an often dramatic decrease
  before the system enters
  the BEC regime.  
\end{abstract}


\maketitle

\section{Introduction}
%

Recently there has been a resurgence of interest in superconductivity
in (quasi-)2D materials. This has been driven by exciting discoveries
of novel superconductors such as magic-angle twisted bilayer graphene
\cite{Cao}, FeSe monolayers \cite{Xue,Hoffman} and transition metal
dichalcogenides \cite{Ye,CastroNeto,Kim}. Many of these and other
interesting superconductors
\cite{Cao,Bozovic2016,Kasahara16309,Cheng2015} appear to belong to the
more strongly correlated class which is distinct from BCS-Eliashberg
superconductors and can be argued \cite{Uemura} to be intermediate
between the BCS and Bose-Einstein condensation (BEC)
limits. 
The challenge then is to develop an understanding of strongly
correlated superconductivity in two dimensions where the long-range
superconducting instability is replaced by a
Berezinskii-Kosterlitz-Thouless (BKT) transition
\cite{Berezinskii,Kosterlitz}.  Meeting this challenge is essential:
an in-depth understanding of these quasi-2D superconductors, requires
that we abandon the predictions of BCS theory. At issue, also is
whether strict BCS theory is appropriate for computing even the
superfluid stiffness; one might expect that this should be obtained by
including contributions of preformed pairs, not present in BCS theory,
at the BKT transition temperature.

Arriving at this formalism is the goal of this paper which addresses
BKT superconductivity in the presence of strong pairing correlations.
Our attention is on the approach and calculation of $T_{\text{BKT}}$
from the normal state, following the extensive body of work on BKT in
atomic Bose systems \cite{DalibardChap}. This is complementary to the
research which addresses $T_{\text{BKT}}$ from the superfluid side
\cite{Kleinert,Leeuwen,Benfatto,Salasnich,Parish,Salasnich2}.  In a
seminal work \cite{HalperinNelson}, Halperin and Nelson  have used a
fluctuation approach to address the physics of approaching the
transition from above.  We argue in this paper that, in line with
their thinking and with Ref.~\cite{Tsinghua}, the normal state
in question should reflect stronger pairing correlations, particularly
those that lead to a stable, observable ``pseudogap".

We stress here that understanding BKT in fermionic systems is
not as straightforward as in their bosonic counterparts.
Indeed the experimental realization of the BKT model was
established in superfluid helium films \cite{Bishop} many years ago.
There is also a convincing case for the observation of
BKT in atomic Bose gases \cite{DalibardChap}.
The nature of the transition and
whether or not it is present in superconducting films has
been a subject of debate \cite{Kapitulnik,Lobb,Armitage,Benfatto2}.
For this reason it is important to pursue a number of different approaches
which address fermionic BKT. This provides the
underlying motivation for our paper and leads us to study the transition
when approached from the normal phase. We do so following the methodology
introduced for atomic gases \cite{Jochim1,Jochim2,DalibardChap}.

\begin{figure}
\includegraphics[width=0.98\linewidth]
{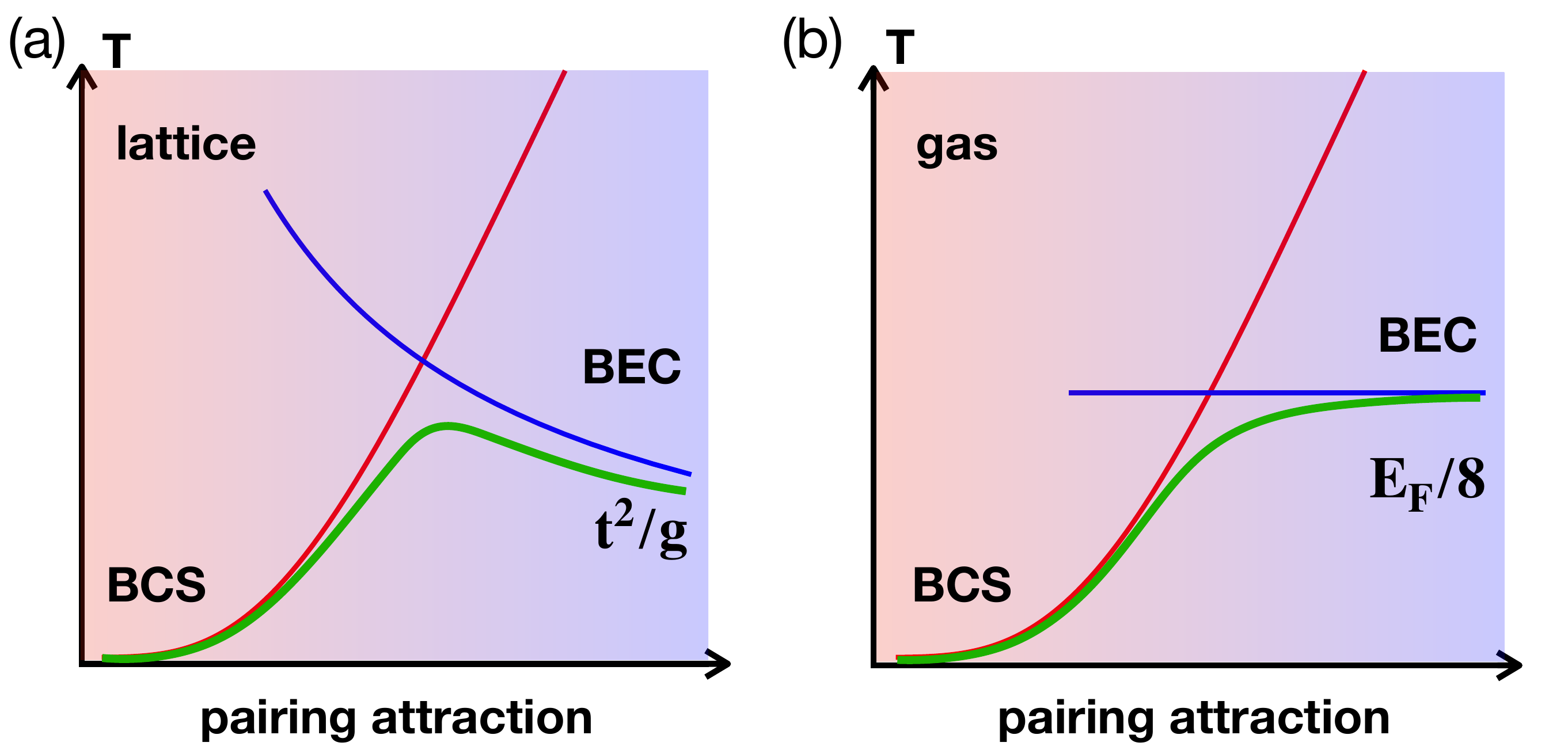}
\caption{Schematic curves showing theoretical expectations from the
  literature for the behavior of $T_{\text{BKT}}$ vs coupling strength
  $g$ for (a) lattice and (b) gas dispersions. The predicted curves are
  embedded inside the two curves labeled ``BCS" and ``BEC". Panel (b)
  can be compared with the data points in Fig~\ref{fig:1}, where some
  differences are evident.  }
\label{fig:0}
\end{figure}

In determining physical variables and consequences, a notable complication is that
plots of $T_{\text{BKT}}$ as a function of the attractive pairing
interaction strength $g$ are non-monotonic, so that knowing
$T_{\text{BKT}}$ does not uniquely determine other fundamental
properties.  Indeed, Quantum Monte Carlo (QMC) simulations
\cite{Scalettar,Nandini} and other more analytic calculations for the
case of a lattice dispersion \cite{Leeuwen}, show plots of
$T_{\text{BKT}}$ vs $g$ which exhibit a superconducting ``dome" shape.
It is generally argued \cite{Scalettar} that this dome lies just
beneath the intersection of two curves: an increasing trend on the BCS
side and a decreasing contribution at larger $g$ representing the BEC
asymptotics, as is shown schematically in Figure~\ref{fig:0}(a).
Similar arguments are presented for the case of a Fermi gas, except
that the BEC limit, rather than decreasing with increasing
coupling constant $g$, reaches a constant asymptote, as shown in
Figure~\ref{fig:0}(b).

A central result of this paper is that when the instability is
approached from the normal state, we, too, find robust domes for the
lattice dispersion, and in addition we find they are present as well
for the case of a gas dispersion.  Importantly, these
non-monotonicities appear in the intermediate coupling regime,
away from the BEC regime.  Here a Fermi surface is still present, as
one would expect in any physical 2D superconducting system.  The dome
arises from a competition between a rising trend of BCS pairing on the
BCS side and the strong suppression of $T_\text{c}$ due to formation of pairs
and the concurrent onset of a pseudogap, well before the BEC regime is
accessed.

The approach in this paper is to combine a pairing fluctuation theory
\cite{ourreview,Chen1,Maly} for BCS-BEC crossover with a description
\cite{Prokofiev,DalibardChap,Phillips} of BKT for bosons, and,
thereby, establish a 2D BCS-BEC crossover theory with a finite
transition temperature.  Because the BKT criterion approaches the
instability from the normal state
\cite{Phillips,DalibardChap,Prokofiev} it reflects the phenomenon of
quasi-condensation or ``presuperfluidity"
~\cite{Jochim2,our2dgas,Cornell}.  From a theoretical perspective
quasi-condensation appears when the bosonic chemical potential becomes
sufficiently small, but non-zero.  This leads to a large number of
non-condensed pairs having very small (but not strictly zero)
momentum.  When approaching the transition from above, it is found
\cite{Phillips,DalibardChap,Prokofiev} that the BKT temperature
depends on the ratio of the effective pair density, $n_\text{B}(T)$,
representing the areal number density of bosons, to their effective
mass, $M_\text{B}(T)$.  The transition occurs when this ratio (which
is proportional to the bosonic phase space density) reaches a critical
value established from previous, (lattice) QMC calculations
\cite{Prokofiev}.

We emphasize that these bosons are a composite made up of fermions and
the fundamental bosonic variables $n_\text{B}$ and $M_\text{B}$ must
depend on the fermionic excitation gap $\Delta(T)$ (which is
non-vanishing even at $T_{\text{BKT}}$).  In a true Bose system
$n_\text{B}$ and $M_\text{B}$ are fixed in temperature.  In fermionic
superfluids both depend on $T$ and interaction strength.  The
non-monotonicity we observe depends on a competition between
$n_\text{B}(T)$, which increases, and $1/M_\text{B}(T)$, which
decreases with increasing $g$.  That the \emph{effective} number of
fermion pairs increases as the attraction becomes stronger should be
clear \footnote{ While all fermions are regarded as paired at zero
  $T$, the residue of the pair propagator (i.e., spectral weight of
  pairs) is small in the BCS regime.}. The pair mass, on the other
hand, can be understood as reflecting the inverse square of the pair
size; the mass is light in strict BCS and becomes heavier with
increased attraction while still remaining within the fermionic
regime.

There is a large body of work which addresses the BKT transition as
approached from below using the Nelson-Kosterlitz
\cite{NelsonKosterlitz} condition. This criterion depends on a
presumed form for the superfluid density $\rho_\text{s}$.  In earlier
calculations it was often assumed that this quantity can be calculated
at the mean field level generalized to include stronger $g$, via a
mean field treatment of the crossover from BCS through BEC.  This
theory presumes that the destruction of the superfluid stiffness
derives entirely from fermionic degrees of freedom.  Application of
this mean field picture for the case of a lattice dispersion
\cite{Leeuwen} or a Fermi gas \cite{Kleinert,Salasnich} generally
leads to plots similar to Figs~\ref{fig:0}(a) and (b).

Clearly, both theoretical approaches (using either the superfluid
density or quasi-condensation) need to be simultaneously pursued by
the community if progress is to be made.  It should be cautioned,
however, that neither of these two schemes explicitly accommodates the
important vortex-antivortex excitations which presumably affect the
size of $T_\text{BKT}$, although how much is not precisely known.
Importantly, in this paper we discuss the relation between the two,
and demonstrate agreement at the weak coupling, BCS level. However, as
the attractive coupling constant $g$ is increased in magnitude,
bosonic excitations become more significant.  Through our comparison
we are able to characterize these bosonic contributions; these also
turn out to be non-monotonic, causing an increase in $T_\text{BKT}$ in
the near ``unitary" regime and a decrease very close to the onset of
the BEC regime.

We stress that our approach which is based on the onset of
quasi-condensation is more directly connected to those experiments
where BKT is most clearly observed as in 2D Bose superfluids
~\cite{DalibardChap,Phillips,Prokofiev} and in 2D Fermi superfluids
\cite{Jochim1,Jochim3} as well.  A quasi-condensation approach
presumes that the inter-boson interactions are sufficiently weak. In
fermionic systems, while there may be strong inter-fermion
interactions, the inter-boson correlations inherent in a BCS-like
ground state are not presumed to be large. Indeed, in the BEC regime,
the inter-pair interaction becomes progressively weaker as the
inter-fermion attraction becomes stronger.

Finally, we end this section by noting that other consequences of
strong pairing correlations should be a central feature of the normal
state.  Indeed, the foundation for using phase only (XY) models in 2D
systems depends on having a substantial pairing at the transition
temperature \cite{Tsinghua}.  Thus, one should characterize a given
superconductor by the pair of temperatures $T_{\text{BKT}}$ and the
pseudogap onset temperature $T^*$, which then removes the ambiguity
associated with the non-monotonicity in the transition temperature.
Pseudogap effects are enhanced in 2D systems and have been clearly
observed in 2D atomic Fermi gases \cite{Kohl,Jochim3}.  For a
transition temperature of, for example $0.08 E_F$, which is rather
strong coupling, we find that the pseudogap onset temperature is
about
twice this temperature.

\subsection{Outline}

We now present an outline of the remaining sections of this paper.
Section II of the paper presents a brief review of our BCS-BEC
crossover theory based on a self-consistent $T$-matrix
approximation. The goal of this discussion is to show how to obtain
the important bosonic quantities $n_\text{B} $ and $M_\text{B}$ for
general $g$ from their fermionic counterparts.

In Section III we discuss the case of 2D superfluids and present an
expression for the BKT transition temperature which is widely used in
the bosonic literature \cite{Phillips,DalibardChap}.  As in
Refs.~\onlinecite{Jochim1,Jochim2}, we show how to apply it to
fermionic superfluids (with both lattice and continuum dispersions).
Contrasting with this ``quasi-condensation" approach to the BKT temperature, is the
more widely used criterion based on the superfluid density, $\rho_\text{s}$, as discussed in Section IV. We present
comparison plots of $T_\text{BKT}$ in the two approaches from above and below the
transition. These are indistinguishable in the weak coupling BCS regime.
However, at moderate or strong coupling, bosonic contributions, which are absent in
the mean-field $\rho_\text{s}$ approach, become increasingly more important.
By comparing these two schemes, we are able to characterize and quantify these bosonic
contributions which are interestingly non-monotonic as a function of increasing $g$.

Reasonable quantitative comparisons with Fermi gas experiments are
presented in Section V, along with predictions for the behavior in the
lattice case.  Comparisons with QMC results on the attractive Hubbard
model indicate some deviation (roughly within a factor of 2).  We show
how to associate the measured transition temperature with other
attendant properties such as the size of the pseudogap.  Following a
discussion in Section VI, our conclusions are presented in Section
VII.

\section{Background }

\subsection{Theory of BCS-BEC crossover}

The incorporation of stronger than BCS pairing attraction is an
important component in addressing BKT in superconductors. In 2D the
$s$-wave pairing instability is always accompanied by two-body bound
states \cite{Randeria} suggesting that pairing effects are
amplified. We thus invoke an extended form of BCS theory associated
with a BCS--BEC crossover. This represents an analytic approach to the
attractive Hubbard model.  Just beyond the BCS regime, we show that
the transition temperature for three dimensional systems is to be
associated with a BEC-like condensation of preformed pairs.  In two
dimensions, $T_\text{c} \equiv 0$, but, in a similar way, we argue
quasi-condensation of preformed pairs depends on a very similar set of
parameters which appear in both 2D and 3D: $n_\text{B}$ and
$M_\text{B}$.


Our approach is intimately tied to the observation \cite{Leggett}
that the
BCS ground state wavefunction
\begin{equation}
\Psi_0=\Pi_{\bf k}(\uk+\vk c_{\mathbf{k},\uparrow}^{\dagger} c_{-\mathbf{k},\downarrow}^{\dagger})|0\rangle
\label{eq:1a}
\end{equation}
has a greater applicability than had been appreciated at the time of
its original proposal. Here $\uk$ and $\vk$ are the BCS coherence
factors. As the strength of the attractive pairing interaction $g$
between fermions is increased, this wavefunction is capable of
describing a continuous evolution from BCS-like behavior to a form of
Bose-Einstein condensation.  The BEC regime sets in when the fermionic
chemical potential $\mu$ becomes negative and at this point,
the underlying Fermi
surface disappears.

We extend this ground state to finite temperature following an
approach to BCS theory proposed by Kadanoff and Martin
\cite{KadanoffMartin,ourreview}.
While in strict BCS theory all bosonic degrees of freedom
(fermion pairs) appear only at (and below) the transition as condensed
pairs, with stronger attraction non-condensed pairs are present. They
are accompanied by the existence of a normal state pairing gap or
pseudogap, which is particularly pronounced
for a 2D system.
For the purposes of our BKT analysis what will be
important is to quantify the number of these preformed pairs $n_\text{B}$ and
their effective mass $M_\text{B}$.

To do so we can associate the underlying structure of BCS theory with a
two particle propagator, called $t_\text{pg}$, which is given
by
\begin{equation}
t_\text{pg}^{-1} (i\Omega_n,\mathbf{q})  = -\frac{1}{g}+ T\sum_{i\omega_l,\mathbf{k}}G(i\omega_l,\mathbf{k})G_0(i\Omega_n-i\omega_l,\mathbf{q}-\mathbf{k})
\label{eq:1b}
\end{equation}
where $G_0$ is the bare fermion Green's function, and the dressed
Green's function $G$ assumes the BCS form
with the Bogoliubov quasiparticle dispersion
$E_{\mathbf{k}}=\sqrt{\xi_\mathbf{k}^2+\Delta^2}$.
We define $\xi_\mathbf{k} = \epsilon_\mathbf{k} -\mu$. A detailed
derivation of Eq.~(\ref{eq:1b}) can be found in
Ref.~\cite{ChenPhD}.
Analytically continuing Eq.~(\ref{eq:1b}) [for 4-vector
$Q \equiv (\mathbf{q},\Omega) \neq 0$] leads to
\begin{equation}
\begin{split}
 t_\text{pg}^{-1}(Q) =& -\frac{1}{g} + \sum_{\mathbf{k}} \left[ \frac{1-f(E_{\mathbf{k}})-f(\xi_{\mathbf{k-q}})} {E_{\mathbf{k}}+\xi_{\mathbf{k-q}}-\Omega -i 0^+}u_{\mathbf{k}}^2 \right.\\
 &\left.  -\frac{f(E_{\mathbf{k}})-f(\xi_{\mathbf{k-q}})}{E_{\mathbf{k}}-\xi_{\mathbf{k-q}}+
  \Omega +i 0^+}v_{\mathbf{k}}^2 \right].
\label{chi_expr}
\end{split}
\end{equation}

Then, as a generalization of the usual Thouless condition, we recognize that the statement $$t_\text{pg}^{-1}(0)=0$$ effectively leads
to the usual BCS temperature dependent gap equation for $\Delta(T)$.
We can think of
this familiar gap equation
as a BEC condition that the pairs associated with $t_\text{pg}$
(which are necessarily non-condensed)
have zero chemical potential $\mu_{\text{pair}}=0$ for \textit{all}
$T \leq T_\text{c}$.


For the most part our interest will be on the long wavelength and low
frequency limit, where the pair propagator can be approximated as
\begin{equation}
 t_\text{pg}(Q) \approx \frac{a_0^{-1}}{\Omega-\frac{\mathbf{q}^2}{2M_\text{B}}+\mu_{\text{pair}}+i\gamma}.\label{eq:pair_prop}
\end{equation}
Here $a_0^{-1}$ characterizes the pair fluctuation strength, $\gamma$ is the
decay rate due to the two-fermion continuum, and $M_\text{B}$ is the
effective pair mass. Both $a_0$ and $M_\text{B}$ can be determined via Taylor
expansion of $t^{-1}_\text{pg}(Q)$ in $\Omega$ and $\mathbf{q}$. In
particular,
$a_0 \Delta^{2} = [n/2 - \sum_\mathbf{k} f(\xi_\mathbf{k})]$ \footnote{Expressions for $M_\text{B}$ in various situations can be found in Ref.~\cite{ChenPhD}.}.
In the presence of a quasiparticle excitation gap, the pair decay rate
at low frequencies vanishes. It is small compared to
$\Omega_\mathbf{q} = q^2/(2M_\text{B})$ when finite momentum pairs make a
significant contribution to the self energy (away from the BCS limit).
From now on we omit $i\gamma$ in the pair propagator.  We
should note that the pair mass $M_\text{B}$ is now accessible through
Eq.~(\ref{eq:pair_prop}).

Next we focus on these non-condensed pairs in the normal state
\cite{ourreview,Chen1}, where the pairs have non-zero chemical
potential $\mu_\text{pair}$ which smoothly vanishes at the transition
into the ordered phase.  Here we identify the pairing gap $\Delta$
with the pseudogap so that $\Delta \equiv \Delta_\text{pg}$. This
excitation gap is to be distinguished from the order parameter.
The self consistency condition can be written as
$t_\text{pg}^{-1}(0)= a_0 \mu_\text{pair}$.  In two dimensions
$\mu_\text{pair}(T)$ will be shown to assume small values, but never
reach zero, except at $T=0$. By contrast, in three dimensions
$\mu_\text{pair}(T)$ vanishes at and below a finite $T_\text{c}$.

To obtain $n_\text{B}$ we note that the self energy associated with
the dressed Green's function is more completely given by
\begin{eqnarray}
    \Sigma(i\omega_l,\mathbf{k}) & =&T\sum_{i\Omega_n,\mathbf{q}}t_\text{pg}(i\Omega_n,\mathbf{q})G_0(-i\omega_l+i\Omega_n,-\mathbf{k}+\mathbf{q})\nonumber\\
    & \approx& -\Delta^2G_0(-i\omega_l,-\mathbf{k}),
\label{eq:4}
\end{eqnarray}
where in this last step we have assumed that the system is near an
instability where $t_\text{pg}(Q)$ is strongly peaked at
$Q=0$. Eq.~(\ref{eq:4}) is a standard approximation in the cuprate
literature for the pseudogap-related self-energy \cite{Maly,Norman}.

We stress that this second line in Eq.~(\ref{eq:4}) is the only
approximation used here, aside from the overarching assumption
implicit in Eq.~(\ref{eq:1a}) that we are dealing with a BCS-like gap
equation and ground state, importantly extended to BCS--BEC
crossover. Note that this approximation effectively ignores Hartree as
well as incoherent contributions to the fermionic self-energy, which
may, as well, introduce particle-hole asymmetry effects. There is an
additional complication (for the lattice case) near half filling
associated with competing charge density wave order in the
particle-hole channel \cite{Scalettar}. For simplicity, we ignore this
here. It is, however, advantageous to adopt the approximation in
Eq.~(\ref{eq:4}) in the vicinity of small $\mu_\text{pair}$ (which is
appropriate near $T_{\text{BKT}}$) for analytical tractability.

Combined with the parametrization in Eq.~(\ref{eq:pair_prop}), we
derive the following self-consistent equations for a fixed-density
system consisting of fermionic and bosonic quasi-particles:
\begin{align}
  a_{0}\mu_{{\rm pair}} &=-\frac{1}{g}+\sum_{\mathbf{k}}\left[\frac{1-2f\left(E_{\mathbf{k}}\right)}{2E_{\mathbf{k}}}
\right], \label{eq:GP}\\
\label{eq:NB}
n_\text{B}&=
                  \sum_{\mathbf{q}}b\left(\frac{q^{2}}{2M_\text{B}}-\mu_{{\rm pair}}\right) = a_{0}\Delta^{2}
                  ,\\
  n&=\sum_{\mathbf{k}}\left[1-\frac{\xi_{\mathbf{k}}}{E_{\mathbf{k}}}\left(1-2f\left(E_{\mathbf{k}}\right)\right)\right]\label{eq:NF},
\end{align}
where $b(x)$ is the Bose-Einstein distribution function. Here $a_0$
and $M_\text{B}$ depend on the three parameters $\mu,\Delta,T$, and
can be deduced through Taylor expansions.

For a 2D system, these equations can be solved self-consistently for
$(\Delta, \mu, \mu_\text{pair})$ at low $T$ and for given interaction
strength $g$. The zero $T$ solution can be taken as the limit of
$T\rightarrow 0$ so that $\mu_\text{pair}$ remains finite in
Eq.~(\ref{eq:NB}) \footnote{Alternatively, at very low $T$, where
  $\mu_\text{pair}$ is very small, one can set it to zero in
  Eq.~(\ref{eq:GP}) so that Eqs.~(\ref{eq:GP}) and (\ref{eq:NF})
  reduce to the BCS-Leggett mean-field equations \cite{Leggett}, which
  can be solved for $(\mu, \Delta)$.  This then also fixes the value
  of $a_0$ and $M_\text{B}$ and thus $n_\text{B}$. Finally, one
  determines $\mu_\text{pair}$ as a function of (low) $T$ via
  Eq.~(\ref{eq:NB}). While this alternative procedure is an
  approximation at nonzero $T$, it becomes exact in the
  $T\rightarrow 0$ limit, where $\mu_\text{pair}$ necessarily
  vanishes. Therefore, we conclude that at $T=0$, the pair density
  $n_\text{B}$ is completely determined by the mean-field solution of
  the ground state, (and so is $M_\text{B}$).}. \emph{We emphasize
  that both $n_\text{B}$ and $M_\text{B}$ are a function of
  temperature}, and should be determined self-consistently via
Eqs.~(\ref{eq:GP}-\ref{eq:NF}) when solving for $T_\text{BKT}$.

What should be clear from this analysis is that in BCS-BEC crossover
theory the normal state consists effectively of an admixture of fermions (with
number $n-2n_B$ and chemical potential $\mu$) and bosons (with number
$n_\text{B}$ and chemical potential $\mu_{\text{pair}}$).
Equation (\ref{eq:NB}) appears physically reasonable in establishing the
direct correspondence between the number of bosons and the energy
scale
$\Delta$ for binding fermions.

\subsection{Behavior of the 3D transition temperature: Hints about
2D}

It is useful to present a few analytic results
from this formalism.
The 3D transition temperature for a gas dispersion is associated with
the condition  $\mu_\text{pair}=0$ at $T_\text{c}$. This enters in the boson number
equation Eq.~(\ref{eq:NB}), and after some algebra, leads to
\[
T_\text{c} \approx \frac{2\pi}{M_\text{B}}\left[\frac{n_\text{B}}{\zeta(3/2)}\right]^{2/3}
\propto \frac{n_\text{B}^{2/3}}{M_\text{B}},
\]
where both $n_\text{B}$ and $M_\text{B}$ are temperature dependent and
calculated at $T_\text{c}$.  \textit{In a very compact way this equation
  encapsulates the behavior of BCS-BEC crossover theory, beyond the
  strict weak coupling limit.} It should be viewed as reflecting the
condensation temperature of preformed pairs.  Importantly, these
represent the emergent bosons which are central to a treatment of
BCS-BEC crossover.  We note that the dependence on $n_\text{B}$ is
similar to what is found in an ideal Bose gas, but it should be
stressed here that inter-boson interactions are present, as is
reflected in the superfluidity \cite{Chen2} and in the collective
modes \cite{Kosztin,Kosztin98} of BCS-BEC systems.  Inter-boson
interactions are associated with both the pairing interaction and the
Pauli repulsion of the underlying fermionic constituents in the pairs.

If, instead, one considers a 2D system, by analogy the associated
number to mass ratio which determines the transition, the transition
temperature might be expected to be $n_\text{B}/M_\text{B}$ where
$n_\text{B}(T)$, represents now the \textit{areal} number density of
bosons, and $M_\text{B}(T)$, their effective mass.  We show in the
next section that this same ratio (known as the phase space density)
appears in the BKT criterion applied by the atomic Bose gas community
\cite{DalibardChap}. Here, however, the bosonic variables are
temperature dependent and depend on the fermionic excitation gap
$\Delta(T)$ and chemical potential $\mu$.

This ratio $n_\text{B}/M_\text{B}$ and its 3D analogue determine the
shape of the transition curves as a function of $g$.  Indeed, the
fractional power $n_\text{B}^{2/3}$ is not very different
quantitatively from $n_\text{B}$, away from the
$n_\text{B}\rightarrow 0$ limit.  In this way, we will see that the
shape of the curves in 2D BKT are not too dissimilar from their 3D
counterparts.
%
%

\begin{figure}
\includegraphics[width=0.9\linewidth]{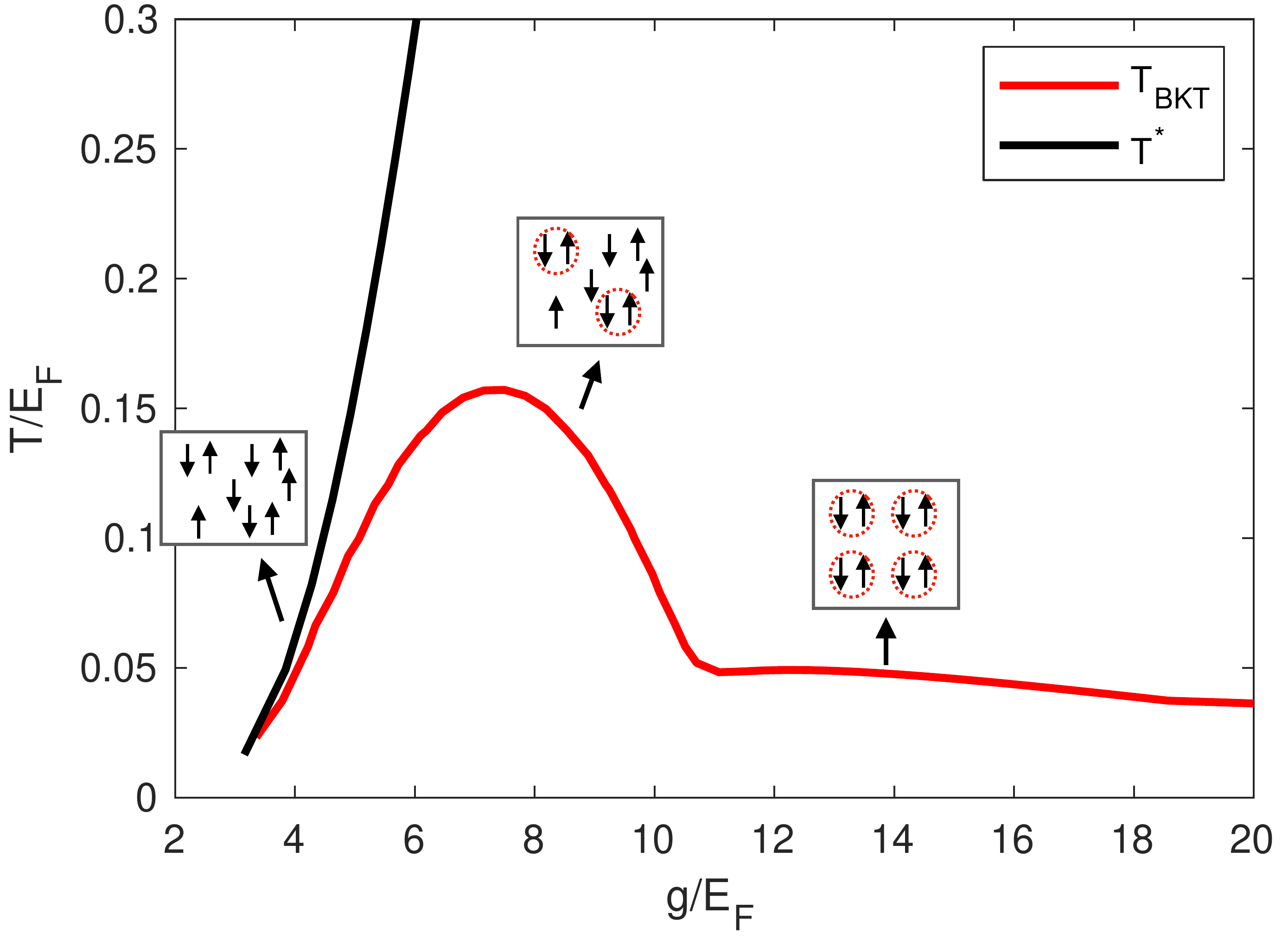}
\caption{The nature of the (normal) state as one varies the strength
  of the pairing interaction and traverses the (schematic, red) BKT
  transition curve.  The black curve indicates the pairing onset
  temperature $T^*$.  This figure serves to define pictorially what is
  meant by the number of bosons $n_\text{B}$.  The square boxes
  represent the relative admixture of fermions and fermion pairs
  (bosons) at the onset of the transition for several values of $g$.
  The bosons are indicated by the paired spins surrounded by dashed
  lines.  The unpaired fermions are indicated by single spins.  The
  bosons (preformed pairs) become more numerous as the interaction
  strength $g$ increases. }
\label{fig:5}
\end{figure}

To elucidate the physics, it is useful to present in Fig.~\ref{fig:5}
an anticipatory plot of $T_\text{BKT}$ as a function of coupling
constant in a way which serves to identify the boson and fermion
constituents. What we indicate in Fig.~\ref{fig:5} is the relative
admixture of broken pairs (fermions) and pairs (bosons) as the
interaction strength is continuously varied.  The small boxes in
Fig.~\ref{fig:5} should be viewed as representative ``cartoons" which
characterize this pseudogap phase.  A very small transition
temperature is expected when the boson number is almost zero, as shown
in the low $g$ regime.  The largest $T_\text{BKT}$ is found in an
intermediate state consisting of bosonic and fermionic
quasi-particles. At the highest value for $g$, all signs of the
fermionic constituents are gone and the transition begins to approach
zero as $t^2/g$.  The pseudogap is present whenever there are a finite
number of pairs at the transition temperature; it becomes
progressively larger, the larger the number of pairs.

\section{BKT Criterion as Approached from the normal state: BKT in Atomic gases}

In the next two sections we discuss two types of criteria which have
been used to establish $T_\text{BKT}$ and follow this with a
comparison.  It is useful to consider Eq.~(\ref{eq:NB}) next for the
strictly two dimensional case where there is no true condensate, away
from the ground state. This equation can be inverted exactly to give
the pair chemical potential:
\begin{equation}
\mu_{\text{pair}} = T\ln\left(1-e^{-n_\text{B}
    \lambda_\text{B}^2}\right) \label{eq:MUP} = T \ln \left(1 - e^{-\mathcal{D}_\text{B}}\right).
\end{equation}
The size of $|\mu_\text{pair}|$, which measures how close the normal
fluid is to a long range-ordered superfluid phase, reflects the
bosonic phase-space density:
$\mathcal{D}_\text{B} (T)
\equiv n_\text{B}(T) \lambda_\text{B}^2$, where $\lambda_\text{B} \equiv \sqrt{2\pi/M_\text{B}T}$ is the de Broglie thermal
wavelength for the 
pairs.
In this notation $k_\text{B} = 1$ and $\hbar=1$.  Importantly, in two
dimensions, $\mathcal{D}_\text{B}$ determines the pair chemical
potential $\mu_{\text{pair}}$, so that there is quasi-condensation
\cite{Jochim2,our2dgas} when $\mathcal{D}_\text{B}$ is sufficiently
large or $|\mu_\text{pair}|$ is sufficiently small.

When approached from the high temperature side
\cite{DalibardChap,Dalibard}, the bosonic BKT transition is known to
occur \cite{Prokofiev} when the temperature dependent phase space
density reaches a critical value
\begin{equation}
\mathcal{D}_\text{B}^\text{crit} \equiv \mathcal{D}_\text{B}(T_\text{BKT})
= \ln (C/\tilde{g})
\label{eq:10}
\end{equation}
where the dimensionless coupling constant $\tilde{g}$ reflects the
size of the 3D inter-boson scattering length $a_\text{B}$, along with
the 2D localization length.  The constant $C \approx 380$ has been
established by QMC \cite{Prokofiev}, based on a tight binding lattice,
but quite generally argued to be universal.  If one parameterizes the
2D confinement by a trap of frequency $\omega_0$, it follows that
\cite{Phillips}
$\tilde{g} = a_B \sqrt{ 8 \pi M_\text{B} \omega_0 / \hbar}$.

Estimates of $\mathcal{D}_\text{B}^\text{crit}$ for fermionic
superfluids are available in the literature
\cite{Jochim1,Jochim2}. The values range from around $4.9$ to $6.45$.
Here we use the value, $\mathcal{D}_\text{B}^\text{crit}= 4.9$, which
best fits the data on Fermi gases \cite{Jochim1}.  This can be
compared with the counterparts in atomic Bose gases which are
typically \cite{Cornell} around $8$.  These are not order of magnitude
variations and the uncertainty does not significantly affect the shape
of the curves for $T_\text{BKT}$ vs $g$; however, it does affect
somewhat their position on the vertical axis.

\begin{figure}
\includegraphics[width=0.8\linewidth]
{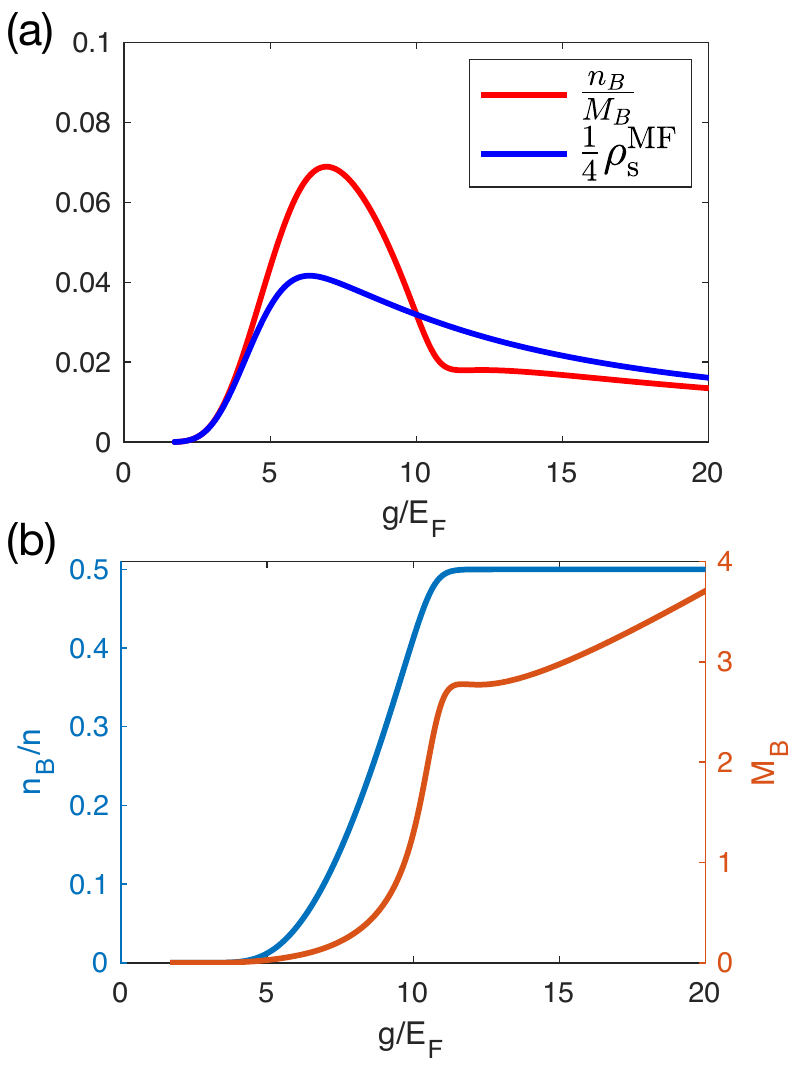}
\caption{(a) Comparison of the ratios $n_\text{B}/M_\text{B}$ and
  $\frac{1}{4}\rho_\text{s}^\text{MF}$ (without
  $\mathcal{D}_\text{B}^\text{crit}$) in the two approaches for the
  lattice case for $n=0.1$.  The
  temperature used throughout this figure is taken to be the critical
  BKT temperature for the $\frac{1}{4}\rho_\text{s}^\text{MF}$ calculation.  The
  difference between the two curves reflects bosonic contributions to
  the destruction of superfluid phase stiffness which are not present in the mean field approach.  Panel (b) shows the two components in the (more
  bosonic) quasi-condensation picture $n_\text{B}$ and $M_\text{B}$.
  This figure indicates that it is a suppression of the pair mass at
  moderate interaction strength which leads to an enhanced maximum in
  the ratios, plotted above. Here $E_\text{F}$ is taken to be the
  non-interacting Fermi energy, with $E_\text{F} \approx 0.604t$, and we take
  the lattice constant to be unity.}
\label{fig:2c}
\end{figure}

\begin{figure}
\includegraphics[width=0.8\linewidth]{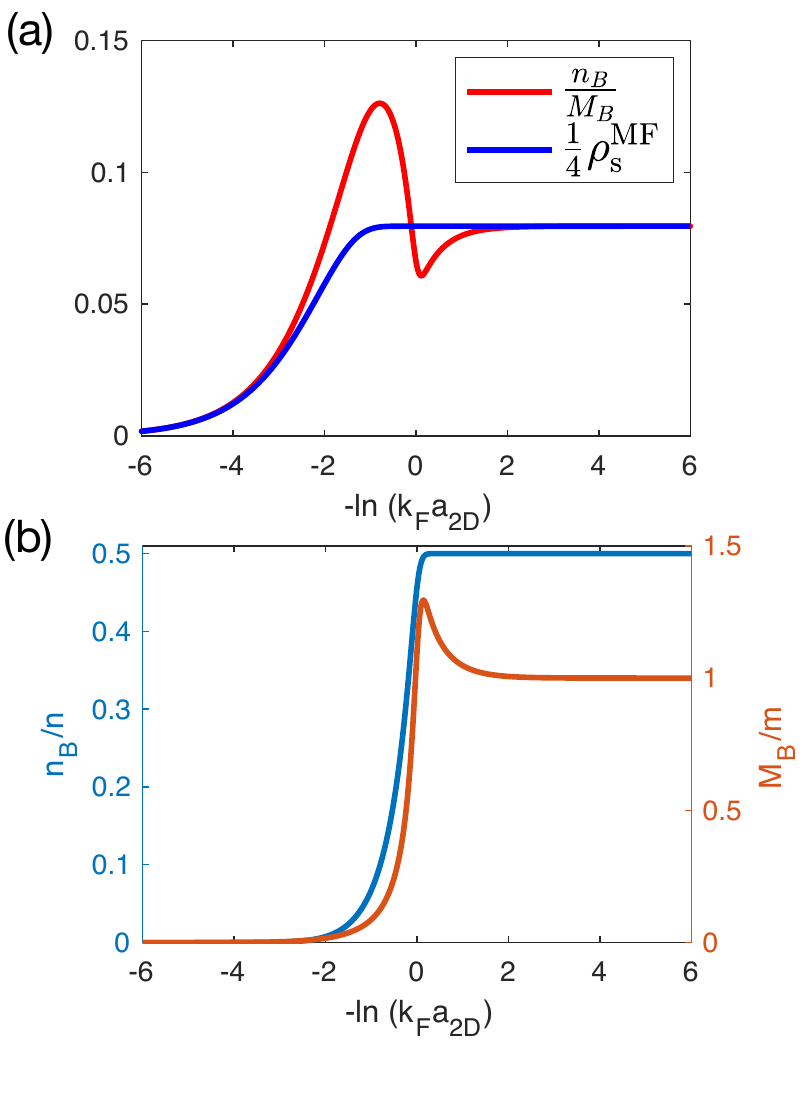}
\caption{(a) Comparison of the ratios $n_\text{B}/M_\text{B}$ and
  $\frac{1}{4}\rho_\text{s}^\text{MF}$ in the two approaches for the gas case.
  The temperature used throughout this figure is taken to be the
  critical BKT temperature for the $\frac{1}{4}\rho_\text{s}^\text{MF}$
  calculation.  The difference between the two curves reflects bosonic
  contributions to the destruction of stiffness, absent in the mean field approach.  Panel (b) shows the two components $n_\text{B}$ and
  $M_\text{B}$.  This figure indicates that it is a suppression of the
  pair mass at moderate interaction strength which leads to an
  enhanced maximum in the ratios.}
\label{fig:2d}
\end{figure}

\begin{figure*}
\includegraphics[width=0.96\linewidth]{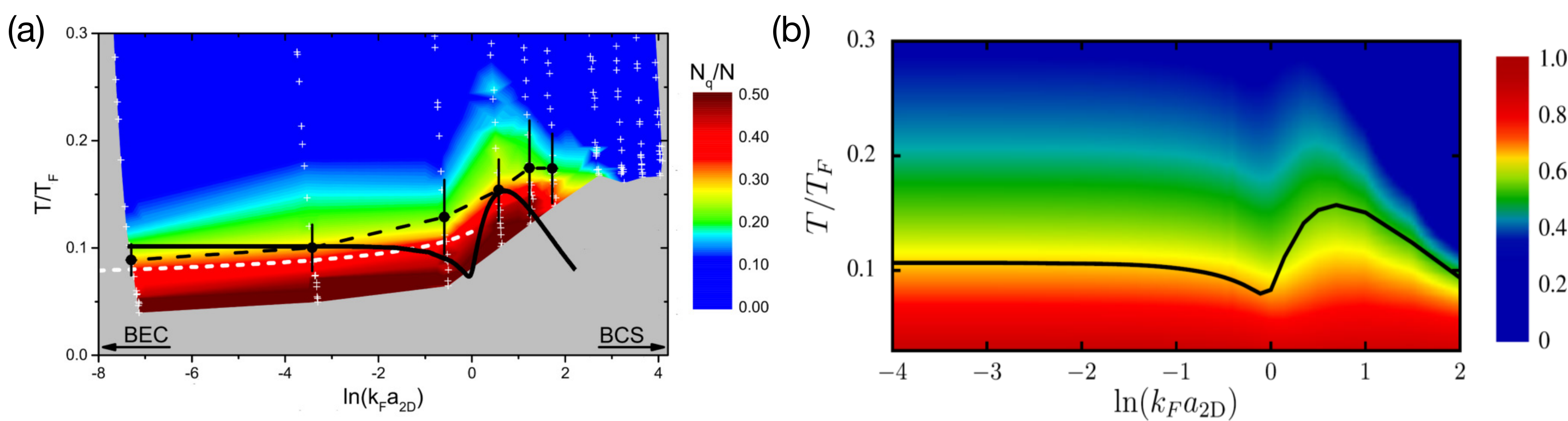}
\caption{(a) Overlay of present theory and experiment for
  $T_{\text{BKT}}$ versus scattering length $a_\text{2D}$ in a Fermi gas.
   The color variations indicate the measured quasi-condensate fractions
  \cite{Jochim1}.
  (b) is from Ref.~\cite{our2dgas}, representing similar
  calculations with a trap included. Here the color variations
  also represent the calculated condensation fractions.}
\label{fig:1}
\end{figure*}

Thus, based on the atomic Bose \cite{DalibardChap,Cornell,Phillips}
and Fermi gas literature \cite{Jochim2}  we apply the BKT criterion
\begin{equation}
  \frac{4}{\mathcal{D}_\text{B}^\text{crit} } \frac{n_\text{B}(T)} { M_\text{B}(T)} = \frac{2T}{\pi}
\label{eq:10b}
\end{equation}
at $T = T_\text{BKT}$.
We stress that $n_\text{B}$ and $M_\text{B}$ in the above equation
reflect the fermionic degrees of freedom, through the pairing gap
$\Delta(T)$.

\section{BKT Criterion Derived from Superfluid Density: Nelson-Kosterlitz Condition}

When approached from the low temperature side, the BKT transition
\cite{Benfatto,scalapino93} occurs at a universal value of the bosonic
superfluid density.  The transition temperature can be defined
\cite{NelsonKosterlitz} in terms of the superfluid component of
$\mathcal{D}_\text{B}$ such that
\begin{equation}
\mathcal{D}^s_\text{B} (T_\text{BKT}) = 4\,.
\label{eq:12}
\end{equation}

To tie the two approaches together, we can, however, extract an inequality
\begin{equation}
\mathcal{D}_\text{B}^\text{crit}
> 4\,.
\label{eq:11}
\end{equation}
This reflects the fact that the \textit{total} bosonic phase space density
must exceed its superfluid counterpart, given in Eq.~(\ref{eq:12})
\footnote{
Indeed, via renormalization group analysis and high precision Monte
Carlo simulations, it is shown that the renormalized and the
mean-field based superfluid densities in the vicinity of the
Kosterlitz-Thouless transition point are different
\cite{Prokofiev}. Therefore, one needs to use a different
$\mathcal{D}_\text{B}$ other than 4.0 in the BKT condition if one is
to use the mean-field based superfluid phase stiffness.}.

If we use BCS-BEC mean field theory \cite{Leeuwen} to evaluate the
superfluid density $\rho_\text{s}^\text{MF}$, Eq.~(\ref{eq:12}) is equivalent
to the condition that at $T_\text{BKT}$, the superfluid density
satisfies
\begin{equation}
\frac{1}{4} \rho_\text{s}^\text{MF}(T) = \frac{2T}{\pi}
\label{eq:12b}
\end{equation}

\subsection{Comparison of the Two BKT Criteria}


Of central importance is to compare these two schemes for the BKT
transition temperature obtained when approached from the normal state
using Eq.~(\ref{eq:10b}) or alternatively using Eq.~(\ref{eq:12b}).
The detailed numerical results in this section and the next are based
on Eqs.~(\ref{eq:GP})--(\ref{eq:NF}) along with Eq.~(\ref{eq:10b}).
This comparison is presented in Fig.~\ref{fig:2c} for the case of a
lattice dispersion and in Fig.~\ref{fig:2d} for the Fermi gas case.

Figures~\ref{fig:2c} and ~\ref{fig:2d} plot the effective ``stiffness
ratios" $n_\text{B}/M_\text{B}$ and $\frac{1}{4}\frac{n_\text{s}}{m}$
(without $\mathcal{D}_\text{B}^\text{crit}$) in the two approaches.
The temperature used throughout this figure is taken to be the
critical BKT temperature for the $\frac{1}{4}\frac{n_\text{s}}{m}$
calculation.  For the Fermi gas case, in place of the attractive
coupling constant $g$, we introduce the 2D fermionic scattering length
$a_\text{2D}$ via
$g^{-1}=\sum_{\mathbf{k}}\frac{1}{2\epsilon_\mathbf{k}+\epsilon_\text{B}^{}}$,
where $\epsilon_\mathbf{k}=k^2/2m$ and
$\epsilon_\text{B}^{}=1/ma_\text{2D}^2$.

We see that the agreement is very good in the strict BCS regime, for
small coupling $g$.  This largely derives from the fact that here
$T_\text{BKT}$ is close to the pairing onset temperature $T^*$. Both
theories yield the same $T^*$.  We can refer to Fig \ref{fig:5} to see
that in this regime the two BKT criteria yield equivalent results, as
the only quasi-particles in the normal state are fermionic and both
are associated with the same pairing onset $T^*$.

A central difference is that the quasi-condensation approach leads to
a higher maximum at intermediate coupling and a more dramatic plummet
in $T_\text{BKT}$ beyond the maximum.  A slight kink appears exactly
when the fermionic chemical potential reaches zero and one might
expect this feature as $n_\text{B}$ has to have a discontinuity in
slope.  Here the boson number density is precisely half the fermion
density and all fermions are paired.  Asymptotically, in the BEC
regime, the two curves also coincide as expected \footnote{In the BEC
  regime, one has $n_\text{B} = n/2$, $M_\text{B} = 2m$, and
  $n_\text{s}/m \approx n/m$ at low $T$, so that
  $n_\text{s}/4m = n_\text{B}/M_\text{B}$.}.

Understanding the physical origin of these two principal differences
is particularly important for arriving at a more complete physical
picture of the BKT transition in a fermionic system.  Both of these
effects arise from the bosonic contributions to the phase stiffness
which are missing in the mean field approximation to $\rho_\text{s}$.  We
refer to the lower panels of Fig.~\ref{fig:2c} and \ref{fig:2d} to
help understand this behavior.

Plotted in these lower panels are the two components in the
quasi-condensation picture $n_\text{B}$ and $M_\text{B}$, with a
rescaling for better visibility. This rescaling will not affect the
deduced ratio plot (except for an overall normalization).  The origin
of the important non-monotonic effects in the ratio can now be seen.
We see that $n_\text{B}$ and $M_\text{B}$ rise just beyond the BCS
regime where pseudogap effects associated with meta-stable
non-condensed pairs begin to emerge.  Notably, the pair mass increases
more slowly (in the plots) giving rise to the maximum in the ratio,
which overshoots the $\rho_\text{s}$-based plot \footnote{In the BCS regime
  for the Fermi gas case in Fig.~\ref{fig:2d}, $M_\text{B}$ scales as
  $(k_\text{F}a)^{-2}$, $n_\text{B}$ scales as $(k_\text{F}a)^{-3}$ and then crosses over
  to $(k_\text{F}a)^{-2}$ in the unitary regime. Thus, $n_\text{B}/M_\text{B}$ scales as
  $1/k_\text{F}a$ to 1.}.  The origin of this slower rise in $M_\text{B}$
is important to understand.  It derives from an increased stability of
non-condensed pairs which is associated with the onset of the
pseudogap.  Stabilization arises because the presence of a pseudogap
means that there is an energy cost, inhibiting the dissociation of
pairs (into fermions). We can think of $M_\text{B}$ as very roughly
representing the inverse square of the pair size. Hence a smaller pair
mass reflects an increased coherence length of pairs.  In this way the
transition temperature exhibits a higher maximum $T_\text{BKT}$.

At increasingly stronger coupling, the bosons contribute a second
structural feature in the $T_\text{BKT}$ plots which appears as a
downturn after the maximum, but before the BEC regime is reached.
This result has been anticipated \cite{Salasnich2}: bosonic
quasi-particles are expected to provide alternative mechanisms for
exciting the condensate. Hence they lead to a reduction in the phase
stiffness and related transition temperature.

\begin{figure*}
\includegraphics[width=0.95\linewidth]{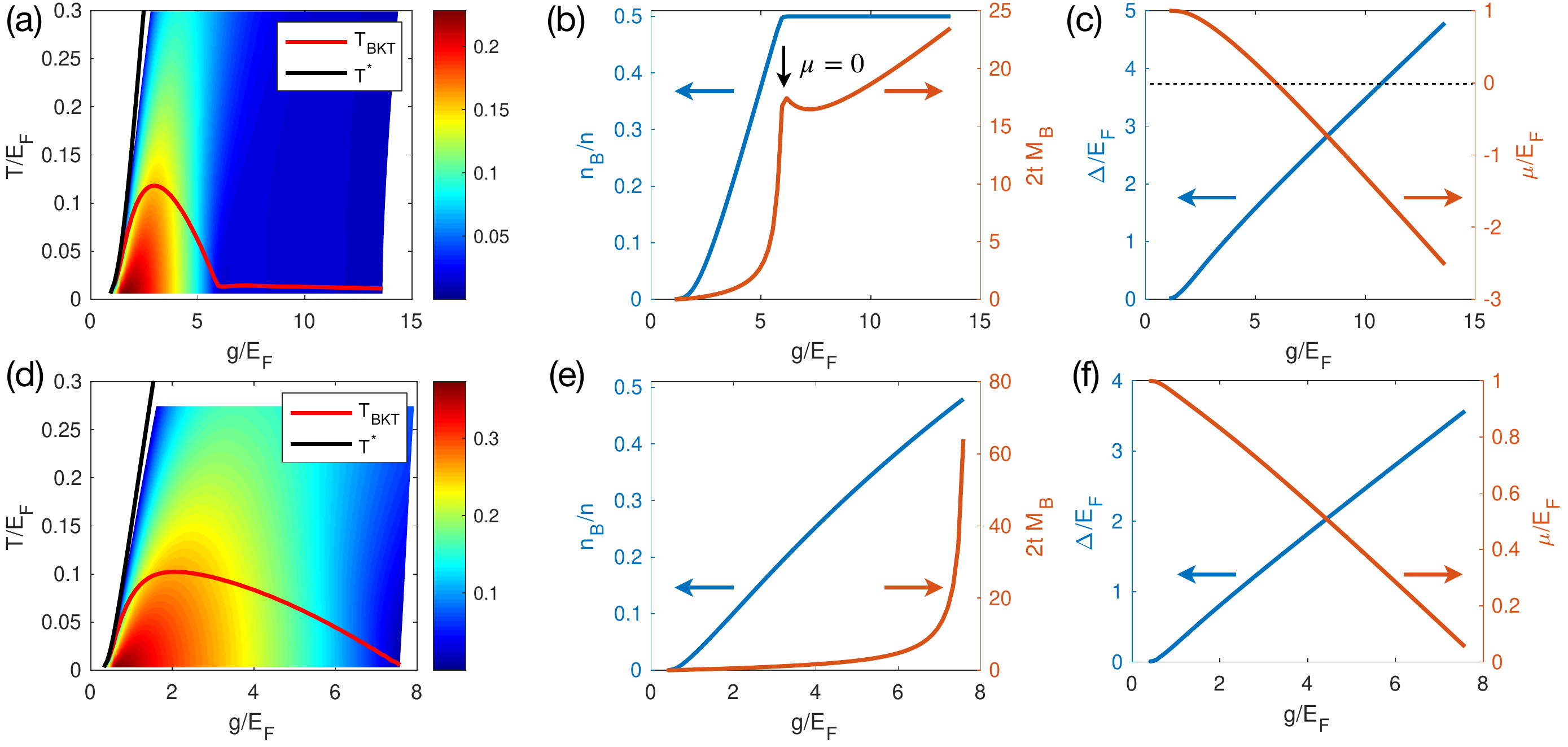}
\caption{BKT temperatures and related parameters for the pair number,
  pair mass, the pairing gap $\Delta$ and the fermionic chemical
  potential $\mu$, for the lattice case. Also indicated is the pairing
  onset temperature $T^*$. Upper panels (a-c) show results for low
  electron density at $n=0.3$, whereas lower panels (d-f) shows
  results for high density at $n=0.7$. Panels (b,c,e,f) are obtained
  at the respective BKT transition temperatures. Note a difference in
  the range of coupling $g/E_\text{F}$ studied for low and high
  electron densities. Here  $E_\text{F}\approx1.67t$ for
  $n=0.3$ , and $E_\text{F}\approx3.28t$ for $n=0.7$. In panel (a),
  asymptotically we observe $M_\text{B} \sim g/t^2$ and
  $n_\text{B} \sim n/2$ as expected in the BEC regime.}
\label{fig:2}
\end{figure*}

We end by summarizing the essential points from this comparison, which
apply to both the lattice and gas dispersion cases. The mean field
$\rho_\text{s}$ approach is missing bosonic contributions \footnote{ Indeed,
  one might imagine that since phase stiffness comes from the quantum
  phase-number duality and the phase is that of pairing field; thus
  the number must be that of pairs as well.  One would then expect
  that phase stiffness must be sensitive to the effective pair
  mass. This mass evidently does not appear in the superfluid density
  in the gas case since $\rho_\text{s}(T=0)=n/m$ is independent of interaction
  effects.}.  In the strict BCS regime these can be neglected and in
that regime the two calculations of $T_\text{BKT}$ are equivalent.
(This equivalence is insured by the particular $T$ matrix used here.)
The most important consequence of including non-condensed pairs in the
gas case is that they lead to a maximum at intermediate coupling. This
derives from the extended stabilization of pairs and concomitant
reduction in their mass.  In the lattice case, for the same general
reason, these pairs enhance an existing (weaker) maximum.

We end this section by noting that correlation functions have also
been addressed \cite{Jochim2,our2dgas} within this quasi-condensation
approach. Theoretically we find that a screened algebraic decay best
fits our numerically obtained results.

\section{Numerical Results}

We can compare our theoretical framework directly to Fermi gas
experiments \cite{Jochim1,Jochim2} on trapped superfluids (although,
in contrast to Ref.~\onlinecite{our2dgas} trap effects have not been
included).  Unlike the curve shown Fig.~\ref{fig:2d} for the mean
field scheme, the experiments at intermediate coupling exhibit a
non-monotonic behavior.  In particular when
$\ln (k_\text{F} a_{2D}) \approx 1$, there is an enhancement of the critical
temperature.  While the extreme BCS limit is not apparent in these
experiments, $T_{\text{BKT}}$ must ultimately reach zero at weak
coupling, so there is a dome like feature \cite{Jochim1} followed by a
nearly constant BEC asymptote.

The left panel of Fig.~\ref{fig:1} presents a direct comparison
between our calculated $T_{\text{BKT}}$ and the experimental data in
units of $E_\text{F}$ versus $-\ln(k_\text{F}a_\text{2D})$, with
$E_\text{F}$ being the non-interacting Fermi energy and $k_\text{F}$
the Fermi wave-vector. From right to left on the horizontal axis
represents the transition from BCS-like to BEC.  The theory curve
(black solid) is overlaid on top of the data showing colored contours
of the quasi-condensate fraction, $N_q/N$. While the data points are
incomplete, with large error bars, in the phase diagram, the overall
agreement between theory and data is reasonably good. The dome
structure in the data for both $T_{\text{BKT}}$ and $N_q/N$ is most
apparent in the edge of the red and green contours, for
$\ln(k_\text{F}a_\text{2D})> -1$. We find a kink near
$\ln(k_\text{F}a_\text{2D})=0$ where the fermionic chemical potential
$\mu = 0$.  Both theory and experiment have to exhibit a decrease in
the transition temperature toward the BCS limit.  Beyond the dip which
establishes the BEC regime, $T_\text{BKT} \approx 0.1
E_\text{F}$. (The calculated asymptotic value is slightly different
from $\frac{1}{8}$ by a factor of $4/4.9$, since the critical value
$\mathcal{D}_\text{B}^\text{crit}$ is slightly larger than 4.)  This
figure is consistent with the expected asymptotic values for
$n_\text{B} = n/2$ and $M_\text{B}=2m$.

We emphasize that by presenting this figure we are not claiming
absolute agreement with experiment.  (Although, perhaps surprisingly,
within error bars, our theory curve passes through all but one data
point with no adjustable parameters).  The experimental figure should
be viewed as a relevant benchmark to help the community arrive at an
understanding of BKT in fermionic superfluids, which is a rather
unique case where there is rather systematic data.  Notably, here we
are dealing with greater complications than, for example, in a
prototypical BKT system such as helium-4.

We replot in the right panel of Fig.~\ref{fig:1} from
Ref.~\cite{our2dgas} the theoretically calculated $T_\text{BKT}$ curve
and the contours showing the quasi-condensate fraction, when trap
effects are included.  Evidently, these trap effects do not
qualitatively affect the general behavior we report above \footnote{As
  a word of caution, it should be noted that the experimental set-up
  for the atomic Fermi gases is only quasi-2D, with a small tunneling
  $t_z$ between neighboring pancakes. Both this
  quasi-two-dimensionality and the trap effect make it possible to
  have a true long range order at low $T$, and they can introduce
  quantitative corrections to $T_\text{BKT}$ as well. To quantify
  these corrections requires sophisticated calculations, beyond the
  scope of the current work. We emphasize that the true long range
  order transition $T_\text{c}$, controlled by $t_z$ and $\omega$, is likely
  much lower than $T_\text{BKT}$, and thus here we ignore its
  influence altogether.}.

We now focus exclusively on the lattice case. Figure~\ref{fig:2}
provides a summary of our results at two representative electron
densities. Panels~(a-c) are characteristic of low electron density
$n=0.3$. As shown in (a), at weak to intermediate couplings,
$T_{\text{BKT}}$ has a dome shape followed by a long slow tail.
Each dome we find is accompanied by a dip
where the chemical potential $\mu$ changes
sign.
The downturn of $T_\text{BKT}$ on the stronger
coupling side of the dome is caused by the increasing contributions of
pairing fluctuations due to increasing pairing strength. The
increasing pairing gap reduces the Fermi level. In addition, these
fluctuations lead to a growing pseudogap at and above $T_\text{BKT}$, which
depletes the density of states and thus suppresses $T_\text{BKT}$. These two
combined effects are so strong that $T_\text{BKT}$ starts to decrease in the
intermediate pairing strength regime, before the Fermi surface shrinks
to zero when $\mu=0$. Beyond this point, the Fermi surface is gone, so
that all fermions are paired up.

Panel (b) shows how the above picture can be regarded as driven by a
competition between an increase in the density of Cooper pairs
$n_\text{B}$ (which saturates to $n/2$ above $g_\text{c}$) and an even
stronger increase in the mass $M_\text{B}$. Here the critical coupling
$g_\text{c}$ is associated with the point where $\mu$ changes sign, as
depicted in panel (c).  For strong coupling $g>g_\text{c}$, the normal state
essentially consists purely of bosonic pairs without unbound fermions
(except at the highest $T$). Note that the pair mass scales linearly
with $g$. This gives rise to the expected asymptotic tail in
$T_{\text{BKT}}\propto t^2/g$.  We emphasize here that the dome at
intermediate couplings is \textit{not} determined by the $t^2/g$
asymptotics seen in strong coupling.  For completeness, in
Fig.~\ref{fig:2}(a) we also present the temperature $T^*$ where the
pseudogap sets in. Over most of the BKT dome, the magnitude of the gap
$\Delta(T_{\text{BKT}})$ at the transition temperature is essentially
unchanged from its zero-temperature value.

Panels~(d-f) are representative results for high electron densities
(here we use $n=0.7$ for illustrative purposes). Just as in the
previous case with $n=0.3$, there is also a superconducting dome in
the range of $g/E_\text{F}\lesssim 8$. In addition, the maximal transition
temperature $T_{\text{BKT}}\sim 0.1 E_\text{F}$ in both cases. However, a
notable difference is that we do not find the long asymptotic tail as it is not
possible to achieve a purely bosonic regime where all electrons bind
into Cooper pairs. This is reflected in the fact that the fermionic
chemical potential (panel (f)) never changes sign before
$T_{\text{BKT}}$ reaches zero. This occurs concurrently with the
vanishing of $\frac{1}{M_\text{B}}$, corresponding to Cooper pair
localization \footnote{Beyond the critical value for $g$, $M_\text{B}$
  changes sign, reflecting a breakdown of the approximations leading
  to Eq.~(\ref{eq:pair_prop})}.

The fact that the fermionic regime is so robust at high densities is
intimately connected to the (near-) particle-hole symmetry of the
underlying lattice Hamiltonian. In a bipartite lattice at exactly
half-filling, the fermionic chemical potential is pinned at $\mu=2dt$
(where $d$ is the dimension), regardless of the interaction strength.
As a result a purely bosonic regime can never be achieved.

Interestingly, within our approach, we observe re-entrant
superconductivity in a narrow range of intermediate electron densities
around $n=0.55$. Here in addition to the dome for $g<g_{c}$, there is
a strong coupling tail with $t^2/g$ asymptotic behavior that sets in
at a slightly larger $g$. Similar re-entrant behavior has
been observed elsewhere \cite{Chenreentrant}.

We can compare to earlier QMC data on the attractive Hubbard model at
$n=0.7$ \cite{Nandini}. There it was found that the BKT transition
temperature reaches a maximum of about $0.175t$ which occurs at
$g = 5t$, as compared with the maximum we find of $0.33t$ which occurs
at $g \approx 6.8 t$. (The $T_\text{BKT}$ calculated using the
mean-field superfluid density yields a maximum of $0.24t$ around $4t$,
also larger than the QMC result). The QMC data do not extend beyond
$g = 8t$. It is likely that the self-energy based approximation
\cite{Maly,Norman} we make as shown in Eq.~(\ref{eq:4}) leads to an
over estimate of particle-hole symmetry and may be in part responsible
for the differences from the QMC data. Additionally, the absence of
particle-hole fluctuations, as in generic $T$-matrix approaches, may
lead to over estimates of the transition temperature and pairing gap
\cite{GMB,ParticleHoleChannel,Petrov2D}. Also important may be
short-ranged charge density wave fluctuations which are neglected in
the present study.

\begin{table}
\centering
{\scalebox{1.1}{
    \setlength{\tabcolsep}{15.0pt}
    \begin{tabular}{cccc}
 \hline\hline\noalign{\smallskip}
$T^*/E_\text{F}$  & $\Delta/E_\text{F}$  & $T^*$ (K)
&$\Delta$ (K)\\[1.5ex]
\hline
0.15  & 0.22 & 2.7 & 4.2\\
0.17 & 0.29 & 3.1 & 5.5\\
\hline
\end{tabular}} }
\caption{Estimates of physical quantities for the case $T_\text{c}/E_\text{F}\approx 0.08$ based on our calculations for $n=0.3$ (in top line) and $ n = 0.7$ in bottom line. To convert to units of temperature,
we assume $T_\text{c} = 1.5$~K. Here $\Delta$ is the pairing gap at the BKT transition. We find
$g/E_\text{F} = 1.87$ and $1.06$ for the low and high densities respectively.}
\label{tab:1}
\end{table}

\section{Discussion}

We turn to Table~\ref{tab:1} for a more quantitative
summary of the various energy and length scales in the
intermediate coupling regime; here for a given ratio of
$T_\text{BKT}/E_\text{F}$, there are two possible values of the
coupling strength $g/E_\text{F}$. For concreteness we choose the ratio
to be 0.08, motivated by estimates made for twisted bilayer graphene
(TBG) \cite{Cao}. We want to firmly stress that this paper
does not incorporate the band structure or other complexities of
this material. (Also note that the maximum transition temperature of the
Monte Carlo calculations \cite{Nandini} does not appear to be
sufficiently large to reach this value.)
\footnote{Nonetheless, we do assume that the superconductivity in TBG
  does come from attractive interactions between electrons, so that
  our general argument for the origin of the dome structure in $T_\text{c}$
  remains valid.}. Nonetheless, as in more conventional BCS theory,
once one knows the transition temperature a number of additional
properties can be quantified regardless of the underlying microscopic
details.

Of particular interest are the size of the pseudogap $\Delta$ at the
transition in comparison to $T_{\text{BKT}}$ and the pairing onset
temperature.  The lower of the two $g$ values appears most reasonable
physically when compared to estimates in TBG \cite{IcRn}.  In both
cases the amplitude of $\Delta$ is relatively the same at
$T_{\text{BKT}}$ and $T=0$; notably, for the smaller $g$, the chemical
potential is close to $E_\text{F}$, so that the system is far from
BEC.  For this more likely situation, we note that the pairing onset
temperature $T^*$ is roughly twice $T_{\text{BKT}}$. When it
differs significantly from
$T_{\text{BKT}}$, this is a
crucially important parameter as it suggests (from
Fig.~\ref{fig:2c}) that this particular material is outside of the
regime where the mean field $\rho_\text{s}$ approach is applicable.
Rather bosonic excitations must be included.

This emphasizes that there are two important temperature scales: $T^*$
and $T_\text{BKT}$.  In general, it is the pair of temperatures
\cite{Tsinghua} which provides full characterization of a given BKT
system.  If it is known that $T_\text{c}/E_\text{F} \approx 0.08$ with
$T_\text{c} \approx 1.5$~K then one can read off from the phase diagram we
present, the size of the pairing gap (around $4-5$~K) and the size of
the pairing onset temperature: ($T^* \approx 3$~K).

\section{Conclusions}

We have stressed that understanding BKT in fermionic systems is
not as straightforward as its bosonic counterpart.
Indeed the experimental realization of the BKT model was
established in superfluid helium films \cite{Bishop} many years ago.
There is also a pretty convincing case for the observation of
BKT in atomic Bose gases \cite{DalibardChap}.
Whether or not this model applies to superconducting films has
been a subject of debate \cite{Kapitulnik,Lobb,Armitage,Benfatto2}.
For this reason it is important to pursue a number of different approaches
for addressing fermionic BKT. We argue that this provides the
underlying motivation for our paper. Here we study the transition
when approached from the normal phase, following the methodology
introduced for atomic gases \cite{Jochim1,Jochim2,DalibardChap}.

An additional motivation for this paper is based on
the excitement behind the recent discoveries of novel 2D
superconductors which appears to be largely based on the hope that these
(often engineered) systems can produce new forms of high temperature
superconductivity. Also exciting is the possibility that they will
serve to teach us about mechanisms for existing high $T_\text{c}$ (say,
cuprate) systems.

Our paper argues for a somewhat more modest perspective.  Independent
of the specifics of the attractive interaction mechanism, in these 2D
systems, there is an absolute maximum to the transition temperature
$T_\text{BKT}$.  It can be rather high, say of the order of $0.1 E_\text{F}$
as found here, or somewhat lower ($0.05 E_\text{F}$ as found in Monte Carlo
\cite{Nandini}), but it does ultimately set an important limit.


\textit{Acknowledgments-} XW and KL were supported by the University of Chicago Materials Research Science and Engineering Center, which is funded by the National Science Foundation under award number DMR-1420709. QC was supported by NSF of China (Grant No.~11774309). We thank Zhiqiang Wang for useful conversations and preparation of a figure.

\bibliography{Review}

\end{document}